\documentstyle[a4,aps,prb,graphics]{revtex}
\begin{document}
  \title{Correlation and Dimerization Effects on the
         Physical Behavior of the  $NR_4 [N\!i(dmit)_2]_2$ 
        Charge Transfer Salts :
        A DMRG Study of the Quarter-Filling t-J Model }
  \author{Marie-Liesse DOUBLET}
  \address{Laboratoire de Structure et Dynamique des Syst\`emes
    Mol\'eculaires et Solides USTL II, B\^at. 15 CC-014, 34095
    Montpellier C\'edex 5, France} 
  \author{Marie-Bernadette LEPETIT} 
  \address{Laboratoire de Physique
    Quantique, IRSAMC, 118 Route de Narbonne, 31062 Toulouse, France} 
  \date{\today} 
  \maketitle

\begin{abstract}

The present work studies the quasi one-dimensional
$N\!i(dmit)_2$-based compounds within a correlated model. More
specifically, we focus our attention on the composed influence of
the electronic dimerization-factor and the repulsion, on the transport
properties and the localization of the electronic density in the
ground-state. Those properties are studied through the
computation of the charge gaps (difference between the ionization
potential and the electro-affinity: IP-EA) and the long- and short-bond
orders of an infinite quarter-filled chain within a $t-J(t,U)$
model. The comparison between the computed gaps and the experimental
activation energy of the semiconductor $NH_2M\!e_2 [N\!i(dmit)_2]_2$
allows us to estimate the on-site electronic repulsion of the
$N\!i(dmit)_2$ molecule to $1.16eV$.

\end{abstract}

\section{Introduction}

Since the discovery of the Bechgaard salts in the
80's~\cite{bechgaard}, the interest for such charge transfer {\em
organic} solids is increasing steadily. The molecular
character of these systems allows a very large diversity in their 
structural arrangement as well as in their chemical composition and 
leads to very rich phase diagrams exhibiting a wide variety of 
attractive physical properties~\cite{review} (charge~\cite{odc,peierls} 
and spin~\cite{ods} density wave phenomena, electronic~\cite{mott} or
structural~\cite{sloc} localizations, metal to insulator~\cite{peierls,mott} 
or metal to
superconductor phases transitions\cite{m-s}, etc \ldots). 
Among the latter, those related to the anisotropy of the conductivity (1D, pseudo-1D or
2D) and hence, to the electronic instabilities of the Fermi 
surface~\cite{review-SF,whangbo,canadell,pouget}
 or to the strongly correlated character of the electronic structure~\cite{strongly}
receive considerable attention. The physical origin of
the conduction band constitutes an important characteristic that
classifies the different compounds into :
\begin{itemize}
\item one-band $3/4$-filled systems for which the conductivity is
insured by the Highest Occupied Molecular Orbital (HOMO) of the donors
---~such as the $TTF$-based molecules~\cite{review,donor}, 
\item one-band $1/4$-filled systems for which the conductivity is
insured by the Lowest Unoccupied Molecular Orbital (LUMO) of the
acceptors ---~such as $M(dmit)_2$ and related
molecules~\cite{acceptor}, 
\item multi-bands systems for which the conductivity is
insured both by the HOMO of the donor molecules and the LUMO of the
acceptor ones~\cite{mixte}.
\end{itemize}

These systems have been extensively and fruitfully studied using non
correlated models such as tight-binding (one-band model) and extended
H\"uckel (orbital model) methods~\cite{tight,eht} for periodic systems. The
topology of the inter-molecular interactions and the occurrence of a 
nesting vector~\cite{pouget} on the Fermi surface is sufficient to understand the 
electronic instabilities that induce first order metal to insulator 
phase transitions~\cite{whangbo,canadell,pouget}, such as Charge Density 
Waves and Peierls distortions~\cite{peierls}. However second order phase transitions, 
that see the metallic properties being gradually destroyed by electron
localization, cannot be explained without the introduction of
electronic repulsion~\cite{mott}. More generally the interaction between the on-site
electron repulsion U, the dimerization strength $\delta$ and the
electronic localization is a crucial problem for second order phase
transitions that requires the treatment of the infinite system within
a correlated model. The problem is further complicated by the
difficulties in determining the numerical values of the
repulsion integrals U. Indeed, these quantities are difficult to obtain
from ab-initio calculations on fragments, since they should take
into account, in an effective way, the effects of the long-range electron
repulsions, whenever those are not explicitly treated in the model. 
This problem is well known since the 50's by users of
semi-empirical methods and an illustrative example is the drop of the
effective Hubbard on-site $\pi$-repulsion of carbon atoms in conjugated molecules
from 11eV for the benzene~\cite{parr} to 6eV for
polyenes~\cite{tchougreff}. An experimental evaluation of the repulsion $U$
is possible from the activation energies of insulating or semiconducting
compounds. However this would require precise theoretical evaluations of the
charge gaps (ionization potential minus electro-affinity) as a
function of $U$ and the structural parameters. 

The physical properties of these charge transfer salts can be modelized
within the Hubbard~\cite{hubbard} or $t-J$~\cite{tJ} formalisms reduced 
to the conduction band.
The resolution of even such simple Hamiltonians requires
a method able to deal with infinite systems. The 
Density Matrix Renormalization Group (DMRG) method~\cite{dmrg}
provides such a procedure for one-dimensional problems.
The aim of the present work will be to modelize
acceptor-based one-dimensional charge transfer salts such as the
isostructural systems $[N\!i(dmit)_2]_2 NR_4$ where $NR_4 = NM\!e_4,\;
NHM\!e_3,\; NH_2M\!e_2$\ldots~\cite{inorg-chem}. We will study a
quarter-filled dimerized chain as a function of both the on-site repulsion 
$U$ and the dimerization factor $\delta$. 
We will compute charge gaps and bond orders of this system, within
an effective, $U$-dependent, $t-J$ model.
Properties for donor-based $3/4$-filled systems can easily be deduced 
by particle-hole symmetry. Since the $NH_2M\!e_2 \; [N\!i(dmit)_2]_2$ compound 
is a semiconductor~\cite{inorg-chem} we will be able to estimate the value of $U$ for the 
$N\!i(dmit)_2$ molecule involved in that type of infinite molecular crystal.

In the next section we describe the DMRG method as well as the model
Hamiltonian. Section~\ref{se:res} reports and analyses the results. The
last section will be devoted to conclusions and perspectives.

\section{Model and method}
\subsection{The model}
\label{se:model}

Single-band quarter-filled systems have an average occupation per site
of only half an electron. Therefore, independently of the value of the on-site
repulsion, it can be expected that the bivalent anion states associated to 
$[N\!i(dmit)_2]^{2-}$ sites are very weakly represented in the ground-state
wave-function. In the non-correlated limit (where the weight of 
the dianionic states is known to be maximum) their contribution is only 
$1/16=0.0625$. 
It is therefore reasonable to eliminate explicit 
reference to the di-anions in the effective model and to allow only
the neutral state $|0\rangle$ (no electron on the LUMO),
and the two singly anionic states $|\uparrow\rangle$ and
$|\downarrow\rangle$ (one electron on the LUMO)
as accessible Valence-Bond states on each site. An often encountered
confusion should however be clarified : the absence of the di-anions in
the wave-function is not synonymous of an infinite on-site repulsion,
as usually assumed by extension from the half-filled band case, but is
only the consequence of their low statistical weight in a
weakly-filled band system. In this case, the main physical effect of
the bivalent anion
molecular states is to lower the effective energy of the local
singlets with respect to the local triplets. This is taken
into account within a $t-J$ Hamiltonian, where the role of the $J$
exchange integral provides this effect.  
\begin{eqnarray}
  \label{tJ}
  H_{t-J} &=& t \sum_{<i,j>} \sum_{\sigma} \left(
    a^{\dagger}_{i,\sigma} a_{j,\sigma} + a^{\dagger}_{j,\sigma}
    a_{i,\sigma} \right) - J \sum_{<i,j>}
  (a^{\dagger}_{i,\uparrow}a^{\dagger}_{j,\downarrow} -
    a^{\dagger}_{i,\downarrow}a^{\dagger}_{j,\uparrow})
  (a^{}_{i,\uparrow}a_{j,\downarrow} -
    a_{i,\downarrow}a_{j,\uparrow}) \\ &=& t \sum_{<i,j>}
  \sum_{\sigma} \left( a^{\dagger}_{i,\sigma} a_{j,\sigma} +
    a^{\dagger}_{j,\sigma} a_{i,\sigma} \right) + 2J
  \sum_{<i,j>}\vec{S_i} \vec{S_j} - J/2 \nonumber
\end{eqnarray}
where $a^{\dagger}_{i,\sigma}$ (resp. $a_{i,\sigma}$) is the creation
(resp. annihilation) operator of an electron of spin $\sigma$ on the
molecule $i$, so that
$a^{\dagger}_{i,\uparrow}a^{\dagger}_{j,\downarrow} -
  a^{\dagger}_{i,\downarrow}a^{\dagger}_{j,\uparrow} $
(resp. $a^{}_{i,\uparrow}a_{j,\downarrow} -
    a_{i,\downarrow}a_{j,\uparrow} $) is the creation
(resp. annihilation) operator of a local singlet on the bond
$<i,j>$.  $t$ refers to the hopping integral between
adjacent ($<i,j>$) molecules $i$ and $j$ while $J$ is the effective
exchange between them.

\subsubsection{The effective exchange integral} 

There are different possible ways to extract the value of the
effective $J$ as a function of the hopping and repulsion integrals. We
choose to extract it in a way such as to reproduce the spectroscopy of a
dimer in the Hubbard models. This choice is motivated by the fact that
the model should be valid over the whole range of $U/|t|$, which 
excludes all perturbative techniques.
The ground-state of the Hubbard Hamiltonian on a dimer is a symmetric singlet $^1\Sigma_g$~:
\begin{eqnarray}
\label{eq:theta}
  ^1\Sigma_g &=& \cos{\theta} {|i\bar{j}\rangle - |\bar{i} j \rangle \over
  \sqrt{2} } + \sin{\theta} {|i \bar{i} \rangle + |j \bar{j} \rangle
  \over \sqrt{2} } \\ \nonumber \\
{\rm of \; energy\; } 
  E_{hub}(^1\Sigma_g) &=& {U - \sqrt{U^2 + 16t^2} \over 2} \nonumber
\end{eqnarray}
while the first excited state is a triplet $^3\Sigma_g$ of energy $0$.
Within the $t-J$ model the first excited state is similarily a triplet
of energy $0$ and the ground-state a symmetric
singlet associated to the energy $-2J$.
The equation of the low energy spectrum in the two formulations yields
immediately 
\begin{equation}
  \label{J}
  J = {U - \sqrt{U^2+16t^2} \over 4}
\end{equation}
Let us notice that the effective exchange integral is a function of
the correlation ratio $U/|t|$, going from $t$ in the uncorrelated
limit to $-2t^2/U$ in the strongly correlated one. Even though $J$ is
extracted from the spectroscopy of a dimer, the validity of this
model on infinite chains has been extensively studied in a previous
work~\cite{nous2}. 

\subsubsection{The correlated effective bond order operator}
\label{sse:bo}

It is clear that the elimination of the explicit reference to the
di-anions in the wave-function will obviously affect the values of any
observable one may be interested in. One must therefore derive
(as done above for the $J$) effective observable operators in order 
to take into account the effects of these states.

In this paper we are more specifically interested in the bond order.
It is easy to see that the bond order (over the $<i,j>$ bond) has a
diagonal matrix in the above $^1\Sigma_g$, $^3\Sigma_g$ basis set and
that its value for the triplet is strictly zero, both in the Hubbard
and $t-J$ models. The value of the ground-state bond order in the Hubbard
model is
\begin{equation}
p_{hub}(^1\Sigma_g) = \sin{2\theta} = {4|t| \over \sqrt{U^2+16t^2}} = {2tJ \over t^2 + J^2}
\end{equation}
where $\theta$ is defined in equation~\ref{eq:theta}.
This value will be assign to the singlet ground-state in the $t-J$
representation. It comes an effective bond order operator
\begin{eqnarray}
  \label{bo}
  \hat p &=& {1 \over 2} \sum_{<i,j>} \sum_{\sigma} \left(
    a^{\dagger}_{i,\sigma} a_{j,\sigma} + a^{\dagger}_{j,\sigma}
    a_{i,\sigma} \right)  \nonumber \\ 
    && + {tJ \over t^2+J^2} \sum_{<i,j>}
  (a^{\dagger}_{i,\uparrow}a^{\dagger}_{j,\downarrow} -
    a^{\dagger}_{i,\downarrow}a^{\dagger}_{j,\uparrow} )
  (a^{}_{i,\uparrow}a_{j,\downarrow} -
    a_{i,\downarrow}a_{j,\uparrow}) \nonumber
\end{eqnarray}

\subsection{The DMRG method}
The DMRG is a very powerful method proposed by S. White~\cite{dmrg} a
few years ago for the treatment of one dimensional spin systems. Since
then it has become one of the leading numerical tools for the study
of quasi-1D correlated quantum systems. This success is due to both
its excellent accuracy for systems as large as a few hundred of sites,
and its flexibility in terms of the model (Heisenberg, $t-J$, Hubbard,
Kondo, etc).

\begin{center}
\begin{figure}[x]
\centerline{\resizebox{9cm}{3cm}{\includegraphics{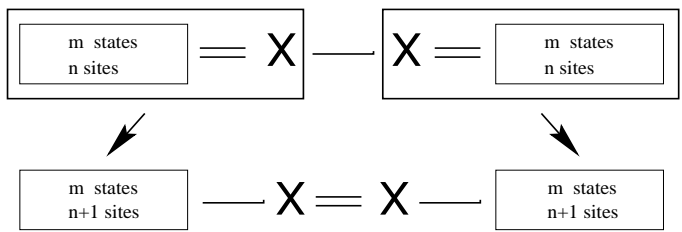}}\vspace{1cm}} 
\caption{ Illustration of the renormalization procedure.  A cross
  represents a $N\!i(dmit)_2$ molecule or $t-J$ site. Single (resp. double)
  lines represent weak (resp. strong) interactions.}
\label{fig:ren}
\end{figure}
\end{center}

In the DMRG approach the properties of the infinite system are
derived by extrapolating the results of a succession of calculations
on finite systems. Each one of these finite-system calculations is
considered as a renormalization group (RG) iteration. The length
$N_{site}$ of the chain increases very slowly at each iteration. For
instance, in the present case, $N_{site}$ is increased by 2 sites (see
Fig.~\ref{fig:ren}).  In spite of the fact that $N_{site}$ increases, the
dimension of the many-body Hilbert space is kept constant by means of
the following procedure illustrated Fig.~\ref{fig:ren}.

At each RG iteration the system is divided in two side blocks spanned
by $m$ local states (typically, $m=60,82 \; {\rm and} \; 100$ in the
present calculations) and two central sites which require $m_1$ local
states, each to be represented exactly (e.g., $m_1 = 4$ for a Hubbard
model and $m_1 = 3$ for a $t-J$ one). The Hilbert space for the
$N^{th}$ iteration (i.e. for the $N_{site} = 2N+2$ system) is obtained
as the set of the antisymmetrized direct products of the four-blocks local
states
\begin{eqnarray*}
  \Phi_{ijkl} = |\phi^1_i \otimes \phi^2_j \otimes \phi^3_k \otimes
  \phi^4_l \otimes \rangle
\end{eqnarray*}
where $\phi^{b}_i$ is the $i^{th}$ local state of the block $b$.
Thus the size of the Hilbert space is kept as $(m \times m_1)^2$ all
along the renormalization scheme.  

Five main steps are involved in going from iteration $N$ to iteration
$N+1$: 
\begin{enumerate}
\item  the ground state of the $N_{site} = 2N+2$ system is obtained
  using a full configurations interaction procedure,
\item the reduced ground-state density matrix of the superblock formed
  by one of the side blocks and its neighboring site is calculated
  (see Fig.~\ref{fig:ren}),
\item the reduced density matrix is diagonalized and the eigenvectors
  yielding the $m$ largest eigenvalues (i.e., occupations) are
  obtained,
\item the superblock local states space of size $m\times m_1$ is projected
  onto the $m$ most populated states derived in the previous step; the
  renormalized interactions within these superblocks, and
  between them and the central sites, are obtained by performing
  the corresponding unitary transformations,
\item the resulting Hamiltonian is then used for calculating the
  ground-state at the following iteration, i.e., we return to the
  first step until convergence is achieved.
\end{enumerate}

The number of RG iterations performed in this work is $100$,
that is 202 effective $N\!i(dmit)_2$ molecules in the chain. The number of block
states kept is $60, 82 \; {\rm and} \; 100$. The infinite system properties are then
extrapolated from those of the finite (but large) systems.

\section{Results and discussion}
\label{se:res}

The $N\!i(dmit)_2$-based systems do not present structural
dimerizations. However, molecular extended H\"uckel (EHT)
calculations exhibit electronic dimerizations through alternated
hopping integrals along the chain~\cite{inorg-chem}. The recent
reasonable description of the transport properties of the
$NH_2M\!e_2[N\!i(dmit)_2]_2$ and the $NHM\!e_3[N\!i(dmit)_2]_2$
systems by Fermi surface EHT calculations\cite{inorg-chem}
has proved the good quality of the calculated electronic
dimerization ratio $\delta = (t_c-t_l)/t_c$ where $t_c$ refers to the {\em
  intra-}dimer hopping integral and $t_l$ to the {\em inter-}dimer
one. $\delta =0$ and $\delta = 1$ correspond respectively to the
non-dimerized and to the product of dimers limits. As announced
previously, we computed the charge gap $\Delta_c$, that is the extrapolated
difference (toward the infinite system limit) between the ionization
potential and the electron affinity as a function of both $U/|t_c|$ and $\delta$.
  If $E(N_{site},N_e)$ is the energy
of a finite $N_{site}$ sites system with $N_e$ electrons : 
\begin{eqnarray*}
\Delta_c &=& \lim_{N_{site}\longrightarrow \infty} \Delta_c (N_{site}) \\ &=& 
\lim_{N_{site}\longrightarrow \infty} \left\{ E(N_{site};N_{site}/2+1) + 
E(N_{site};N_{site}/2-1) - 2*E(N_{site};N_{site}/2)\right\}
\end{eqnarray*}
. $U/|t_c|$ spans the whole range
of correlation strength between $0$ and $+\infty$, while $\delta$ is
varied between the two extreme values given by the real systems, that
is $\delta=0.05$ for the metallic phase $NHM\!e_3[N\!i(dmit)_2]_2$ and
$\delta=0.75$ for the semiconducting one $NH_2M\!e_2[N\!i(dmit)_2]_2$. In
addition to the macroscopic picture given by the charge gap, we aimed
to have a closer, microscopic understanding of the ground-state
wave-function and of its localized versus delocalized character, that
cannot be provided by extended H\"uckel calculations. These properties
can be traced using the chemical bonding on the dimers, and more specifically
the ratio between long- and short-bond order parameters 
\begin{eqnarray*}
\lambda = {p_l \over p_c} = \lim_{N_{site}\longrightarrow \infty}
{p_l(N_{Site}) \over p_c(N_{Site})}
 \end{eqnarray*}
 where $p_c$ and $p_l$ are respectively the short- and long-bond
 orders. $\lambda$ is expected to vary from $0$ for the product of
 dimers limit to $1$ for the totally delocalized system.

Owing to the present renormalization scheme that increases the number of
sites by only 2 at each iteration, one encounters two specificities.
\begin{itemize}
\item The number of electrons is increased by 1 at each
  renormalization iteration, the ground state being a singlet for odd
  iterations ($N=2p-1$, $4p$ sites and $2p$ electrons) and a
  doublet for the even ones ($N=2p$, $4p+2$ sites and $2p+1$)
  electrons. As a consequence all computed observables present an
  even-odd iterations alternation, requiring to compute separately the
  odd and even iterations limits ---~even though they both should
  converge to the same value.
\item In order to keep a physically meaninfull system, one must impose
  the edge bonds to be consistently short. Thus, depending on the parity
  of the iteration, this leads to a long/short alternation of the central bond, 
  on which the bond orders are computed.
\end{itemize} 

 Three series of calculations have been performed for $m=60, \; 82,
 \;{\rm and }\; 100$ block states kept. Let us notice that the
 $m=82,\;{\rm and }\;100$ calculations treat exactly the finite
 systems up to $10$ sites while the $m=60$ only treats them exactly
 up to $8$ sites.

\subsection{Charge Gap}

\begin{center}
\begin{figure}[x]
\resizebox{8cm}{8cm}{\includegraphics{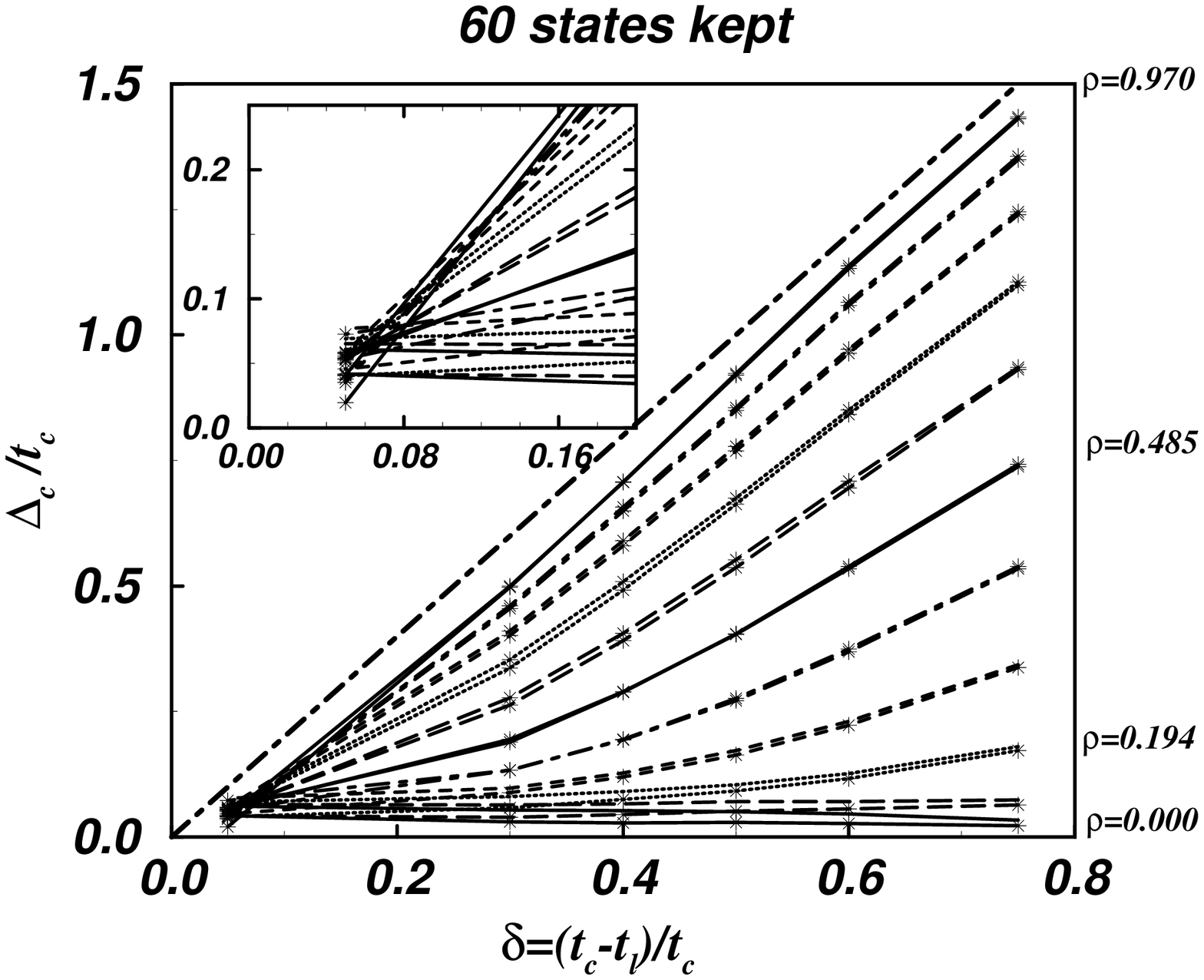}}
\resizebox{8cm}{8cm}{\includegraphics{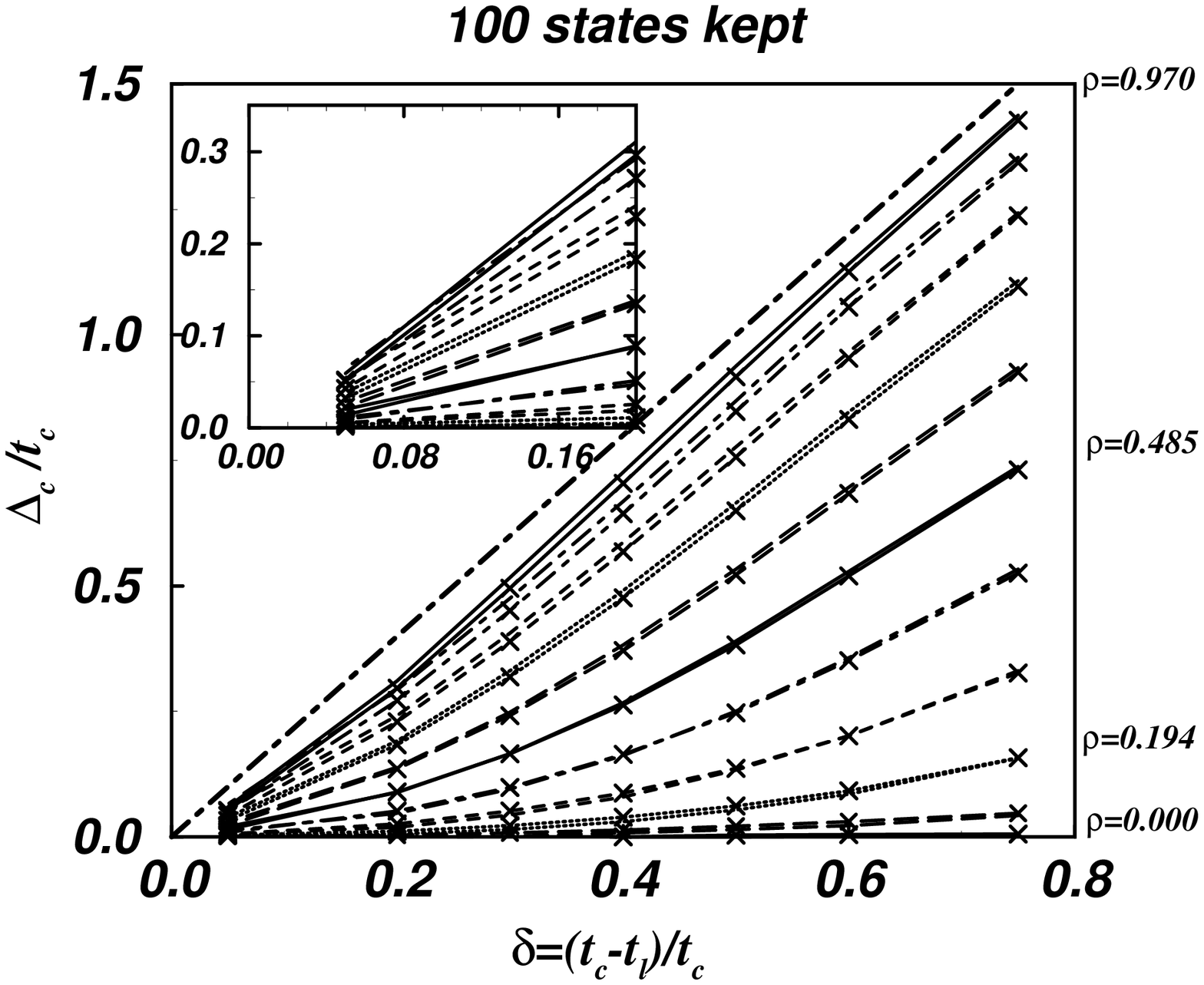}} 
\caption{Charge gaps in $t_c$ units as a function of the dimerization
factor $\delta$. Each curve corresponds to a different value of $J/t$,
therefore of $\rho=U/(U+4t_c)$ which varies between $0$ and $0.97$ by
step of $0.097$.  (a) $m=60$ and (b) $m=100$
block-states are kept.}  
\label{fig:gap}
\end{figure} 
\end{center}

Figures~\ref{fig:gap}a-b present the $\delta$ dependence of the charge
gap for 8 equidistant $U/(U+4t_c)$ values in the range $[0,1]$. They
correspond respectively to $m=60$ and $m=100$ block states kept at
each iteration.  The results for $m=82$ states have not been presented
here since they are not qualitatively ---~and even quantitatively~---
much different from those obtained with $m=100$.  

While the two figures exhibit quite similar results for large
dimerizations and large $U$, they differ subtancially when either $U$
or $\delta$ are small. Indeed, in these latter regions, the $m=60$
calculations largely overestimate the values of the gaps compared to
the ones obtained with $m=100$. In the large dimerization region
---~$\delta > 0.5$~--- the increasing character of the gap as a
function of the correlation strength $U$ is well reproduced,
eventhough the discrepancy with the $m=100$ curves increases rapidly
when $U$ decreases.  In the small dimerization limit however, the
$m=60$ results do not even exhibit the correct behavior as a function
of $U$. Indeed, for $\delta=0.05$, the smallest computed gap corresponds
to the largest value of $U$ (the extrapolated gap for $U=0$ and
$\delta=0$ is $0.05$ instead of $0$). Moreover,  unphysical
crossings between $U$-curves can be observed. Those artefactual
effects can be directly related to an insufficient number of block
states kept in the $m=60$ calculations, since they disappear when $m$ 
is increased to $82$ or $100$. This larger $m$ value required for an
adequate treatment in the small $\delta$ and/or small $U$ region can be
rationalized in the following way. 
\begin{itemize}
\item The populations of the local block states in the total
wave-function are known~\cite{marieB} to follow a decreasing exponential of
the form $\gamma = \exp(-\alpha(U) n)$, where $\alpha(U)$ is a
decreasing function of $U$. As documented in ref~\cite{marieB}, the accuracy
of a DMRG calculation is directly related to the fraction of the total
population provided by the $m$ block-states kept. Hence, as $U$
decreases, a larger number of states $m$ needs to be kept.

\item An optimal use of the DMRG procedure supposes that the
perturbation imposed on the electronic wave-function follows a linear
dependence with the system size. In the present case, the addition or
removal of an electron is a major perturbation for small size systems
($50\%$ of the electronic population in the first iteration for the
$N_{site}/2-1$ system) whereas it is a negligible one for larger
systems ($100/(N+1)\%$ at the $N^{\rm th}$ iteration). This problem
can only be overcome by insuring that all local states needed for the
description of the large systems (weakly perturbed) are not removed in
the first few iterations. That is by increasing the size of the largest system
treated exactly and equivalently enlarging $m$ (8 sites for $m=60$, 10
sites for $m=82$ or $m=100$).  
\end{itemize} 
From now we will focus our attention on the $m=100$ results.

As expected the charge gap is an increasing function of both the
dimerization ratio and the on-site repulsion. In agreement with the
quarter-filled dimerized H\"uckel model, the $U=0$ curve does not
exhibit any significant gap. Its maximum computed value is
$0.005|t_c|$ and can be considered as characteristic of
the calculations precision. Similarly, all curves extrapolate towards a zero gap
for $\delta=0$. 

In large $U$ limit the exchange $J$ parameter tends
to $0$~; the infinite-$U$ Hamiltonian  is therefore limited to the
kinetic part acting on a Hilbert space excluding all doubly occupied
sites. 
\begin{eqnarray*}
H_{U\rightarrow +\infty} = \sum_{<i,j> \;\sigma}\left(
  a^{\dagger}_{i\sigma}a_{j\sigma} +  a^{\dagger}_{j\sigma}a_{i\sigma}\right)
\end{eqnarray*}
The spin part of the latter Hamiltonian is completely uncoupled from
the space part and the system is equivalent to a spinless fermion one.
As shown in ref.~\cite{penc} the ground-state energy per site can be
expressed as
\begin{eqnarray*}
  E = t_c {2-\delta\over \pi} E\left(\phi,2{\sqrt{1-\delta}
        \over2-\delta}\right)  
\end{eqnarray*}
where
\begin{eqnarray*} 
\phi &=& \left\{ \begin{array}{lll} 
 \pi n & for & n>1/2 \\ \pi(1-n) & for & n<1/2 
\end{array} \right.
\end{eqnarray*}
$n$ refers to the band filling, $E$ to the 
elliptic integrals of the second kind. One sees immediatly
that the energy derivative presents a discontinuity as a function of
the band-filling, discontinuity that can be associated to the charge
gap. 
\begin{eqnarray*}
{\Delta_c \over t_c} =  {2-\delta \over \pi} 2 \pi \sqrt{1- \left(2{\sqrt{1-\delta}
        \over2-\delta}\right)^2} =  2\delta
\end{eqnarray*}
Turning back our attention to fig.~\ref{fig:gap}b one sees that our
calculations converge well to the above expected limit for
$U\rightarrow +\infty$.

One of the main interest of having an accurate estimate of the charge
gap within a correlated model is the ability it provides to evaluate
the effective repulsion $U$. Organic conductors are well-known to
behave as strongly correlated systems, since the rather weak
inter-molecular interactions lead to quite large $U/|t|$ ratio.
However the difficulties encountered to obtain reliable $U$ has led to
very large controversies for years~\cite{Uvalues}.  On one hand, a non
biased extraction from experimental datas would require the use of a
correlated interpretative model beyond the standard non or weakly
correlated methods (H\"uckel, RPA~\cite{rpa}). On another hand, direct
computation from ab initio quantum chemistry methods on the molecular
unit is expected to (largely) overestimate the value of $U$.  As
already mentionned in the introduction, the $U$ integral is an
effective integral that takes into account number of other correlation
effects than the purely one-site repulsion.

The availability of accurate gap values within a correlated model,
valid over the whole $U/|t_c|$ range, allows us to confront them to the
experimental activation energies in order to extract reliable values
of $U$. Single-crystal temperature-dependent conductivity
measurements have disclosed an activation energy of 
$E_a = 0.21eV$~\cite{inorg-chem}, 
at room temperature, for the semiconductor
$NH_2M\!e_2 \; [N\!i(dmit)_2]_2$. Considering that at room temperature
the gap is reduced by the thermal energy, the $k_bT$ value should be
added to $E_a$ in order to obtain its $T=0$ estimation.
One therefore obtains an estimated experimental gap at $T=0K$ of
$E_a(0K)= 0.22eV$, that is $E_a(0K)=0.76|t_c|$ (where $t_c=0.29eV$ has
been taken from EHT calculations~\cite{inorg-chem}).  This corresponds
to $U = 4.0|t_c| = 1.16eV$ for $\delta=0.75$. Following the same line
of argument for the $NHMe_3 \; [N\!i(dmit)_2]_2$ compound, the $\frac
{1}{2} k_bT$ factor appears to be larger than the computed gap for the
whole range of $U/|t_c|$. Although the calculated gap $\Delta_c$ never
equals zero, the small values it takes are in agreement with the
metallic behavior exhibited by this system. Using now the value of $U$
obtained for the $N\!i(dmit)_2$ molecule and the EHT $t_c=0.237eV$
value for the $NHMe_3 \; [N\!i(dmit)_2]_2$ system, the correlation
ratio comes to be $U=4.89|t_c|$, slightly larger than for the
semiconducting phase.

\subsection{Bond Order}

\begin{center}
\begin{figure}[x]
\resizebox{8cm}{8cm}{\includegraphics{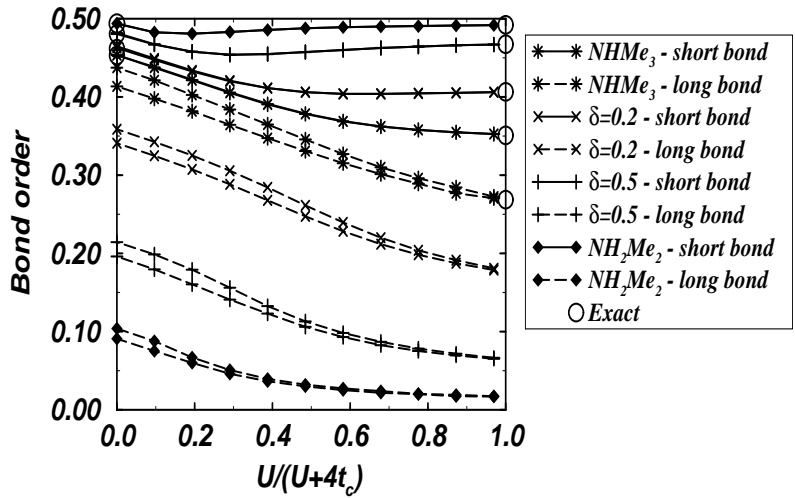}} 
\caption{Long- and short-bond orders as a function of $U/(U+4t_c)$.
Solid line: short-bond order, dashed line: long-bond order. 
The circles correspond to analytical evaluations of the asymptotic
values for $U=0$ and $U \longrightarrow \infty$.}
\label{fig:bo}
\end{figure}
\end{center}

Fig.~\ref{fig:bo} reports the short- and long-bond orders (respectively
referenced as $p_c$ and $p_l$) as a function of $U/|t_c|$ and for
different dimerizations. $p_c$ and $p_l$ have been computed according
to the expression derived in section~\ref{sse:bo}, i.e. the effects of
the di-anionic configurations are taken into account in an effective
way. From a technical point of view, the bond order was computed on
the central bond at each iteration. In the present renormalization
scheme, this bond alternates from long for odd iterations, to short
for even iterations. Additional oscillations occur between $(4p)$- and
$(4p+2)$-electrons systems for long bonds and $(4p+1)$- and $(4p+3)$-electrons
systems for short bonds. We therefore performed 4 independent
extrapolations. For short bonds, the $(4p+1)$- and $(4p+3)$-electrons
extrapolations converge towards the same values all over the $U/|t_c|$
range. However, as can be seen on Fig.~\ref{fig:bo}, for long bonds, the
$(4p+2)$- and $(4p)$-electrons extrapolated $p_l$ curves split over for small
$U$, following the aromatic, anti-aromatic alternation~\cite{aromatic}.

As expected, when the dimerization factor goes to 1, the system
tends toward a product of independent singly-occupied dimers, that is
$p_c$ tends to $0.5$ and $p_l$ to $0$. Looking now at the
$U=0$ and $U\rightarrow +\infty$ limits, one can perform an analytical
calculation as a function of $\delta$. It is easy to reach the
H\"uckel result 
\begin{eqnarray}
p_c(\delta) &=& {1\over \pi} \int_0^{\pi/2} { 1+
(1-\delta)\cos{\theta} \over \sqrt{1 + (1-\delta)^2 +
2(1-\delta)\cos{\theta}} } \; d\theta
\end{eqnarray}
The $U\rightarrow +\infty$ result can be obtained from the spinless fermion case
as treated in ref.~\cite{penc}, and 
\begin{eqnarray}
p_c(\delta) &=& <c^{\dagger}_0 c_1>  \nonumber \\
&=& {1\over 2\pi} \left[ \delta F(1/2,q) + (2-\delta) E(1/2,q) \right]
\end{eqnarray}
where 
\begin{eqnarray*}
E(1/2,q) &=& \int_0^{\pi/2} {1\over \sqrt{1 - 4q\sin^2{\theta}}} \;
d\theta \\
E(1/2,q) &=& \int_0^{\pi/2} \sqrt{1 - 4q\sin^2{\theta}} \;
d\theta \\
q &=& {1-\delta \over (2-\delta)^2}
\end{eqnarray*}
The asymptotic values are represented by open circles in
Fig.~\ref{fig:bo}, all of them in very good agreement with our
computed curves.

\begin{center}
\begin{figure}[x]
\resizebox{8cm}{8cm}{\includegraphics{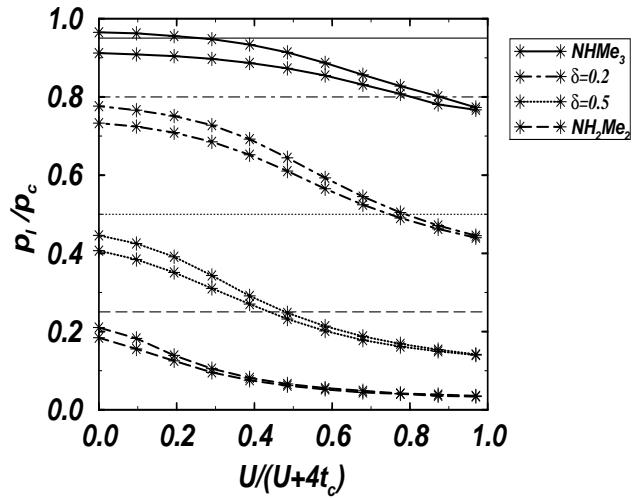}} 
\caption{Bond-orders ratio: $p_l/p_c$.
Solid line: $\delta=0.05$, dash-dotted line: $\delta=0.20$, dotted
line: $\delta=0.50$, long-dashed line: $\delta=0.75$.}
\label{fig:borap}
\end{figure}
\end{center}

According to the dimerization factor $\delta$, the different systems
studied in fig.~\ref{fig:bo} present two types of behavior. The first
type, which will be referred as {\em localized}, includes the
$\delta=0.5$ and $\delta=0.75$ ($NH_2M\!e_2\;[N\!i(dmit)_2]_2$)
systems. It is characterized not only by a small $p_l/p_c$ ratio (as
seen in Fig.~\ref{fig:borap}) but also by a non monotonic behavior of the
short-bond order as a function of $U/|t_c|$. Starting from the H\"uckel
limit, $p_c$ begins to decrease, goes through a minimum
($p_c^{min}(\delta)$), then increases up to a $U\rightarrow +\infty$ limit,
slightly smaller than the $U=0$ one. The $U/|t_c|$ value associated to
$p_c^{min}(\delta)$ is a decreasing function of $\delta$.  The second
type of behavior, to which we will refer as {\em delocalized}, is
represented by the $\delta=0.05$  ($NHM\!e_3 \,
[N\!i(dmit)_2]_2$) system. It is characterized by a $p_l/p_c$ ratio close to
$1$ and a monotonically decreasing $p_c$ curve as $U/|t_c|$ increases.
The $\delta=0.2$ $p_c$ curve however exhibit a intermediate behavior
between the two previous ones, $p_c$ first decreases as $U$
increases, then reaches an asymtptotic value around $U=4|t_c|$.

The long-bond order present a qualitatively (but not quantitatively)
similar behavior for all dimerization values, decreasing as the
correlation strength raises and exhibiting an inflexion point. The
latter occurs for smaller $U/|t_c|$ values as $\delta$ increases. 
As expected the long-bond order diminishes as $\delta$ increases,
while the reverse tendancy is observed for the short-bond order. 

Figure~\ref{fig:borap} presents the $p_l/p_c$ ratio as a function 
of the correlation strength. 
It can be seen in this figure that ---~as expected~--- the
correlation has a localizing effect since $p_l/p_c$ diminishes as $U/|t_c|$
increases for all values of $\delta$.  It is more interesting to
compare the hopping-integral ratio (horizontal lines) to the
bond-order one. The {\em delocalized} system exhibit
similar $t_l/t_c$ and $p_l/p_c$ ratios. On the contrary,  the {\em localized}
systems show much smaller long- to short-bond order ratios than the
hopping-integrals do, i.e the collective effects seem to strongly
enhance the localization of the electron density on the dimers. 

\section{Conclusion}

The present work deals with the correlation and dimerization
effects in quarter-filled quasi one-dimensional organic conductors
such as the $X \; [M(dmit)_2]_2$ radical anions compounds (where $M$
stands for $N\!i$, $P\!t$ or $P\!d$). We treated them as a
correlated dimerized chain within a $t-J$ model. The effective 
exchange integral $J$ is taken as a function of the hopping integral
$t$ and the correlation strength $U$, derived to reproduce the dimer
low energy spectroscopy. We computed accurately the charge gaps for
the infinite chain (using the DMRG method) over the whole range of
electronic repulsion and dimerization factor. The comparison between
our theoretical results and the experimental activation energy on
semiconducting $N\!i(dmit)_2$-based compounds allowed us to extract a
reasonable evaluation of the one-site repulsion integral $U$ for the
$N\!i(dmit)_2$ molecule: $1.16eV$, i.e. $U/|t_c|=4.0$ for
the semiconducting $NH_2M\!e_2\;[N\!i(dmit)_2]_2$ compound, and 
$U/|t_c|=4.89$ for the metallic $NHM\!e_3\;[N\!i(dmit)_2]_2$ one. 
It would be interesting to extract,
in the same way, the $U/|t|$ value for other molecules widely encountered
in such molecular compounds, specifically since there is a large number 
of controversies about their numerical values. 

We also computed the long and short bond order as a function of both
$\delta$ and $U$ and showed that the collective effects enhance,
in strongly dimerized systems, the electronic localization on the 
dimer units.
Meanwhile an unexpected $U/|t_c|$ behavior has be found for
these dimerized systems, with the observation of a minimum on the
short-bond order curves. The increase of $p_c$ for large $U$ support
the idea that the electron repulsion enhance the accumulation of
electrons on the strong dimers, leading to a product of dimer
zeroth-order picture even for moderately dimerized systems ($\delta=0.5$).

\bibliography{ml1_5}

\newpage

\end{document}